\documentclass[twocolumn,amsmath,amssymb]{revtex4}
 
\usepackage{graphicx}
\usepackage{dcolumn}
\usepackage{bm}

\begin{document}
\newcommand{\be}{\begin{equation}}
\newcommand{\ee}{\end{equation}}
\newcommand{\bea}{\begin{eqnarray}}
\newcommand{\eea}{\end{eqnarray}}
\newcommand{\n}{\nonumber\\}

\title{\bf The Backbone of a City}

\author{Salvatore Scellato$^{1}$, Alessio Cardillo$^{2}$,  Vito Latora$^{2}$ 
and Sergio Porta$^{3}$}
\affiliation{$^{1}$ Scuola Superiore di Catania, Italy}
\affiliation{$^{2}$ Dipartimento di Fisica e Astronomia, 
Universit\`a di Catania, and INFN Sezione di Catania, Italy}
\affiliation{$^{3}$  Dipartimento di Progettazione dell'Architettura, 
Politecnico di Milano, Italy}
\date{\today}

\begin{abstract}
Recent studies have revealed the importance of centrality measures 
to analyze various spatial factors affecting human life in cities. 
Here we show how it is possible to extract the backbone of a city 
by deriving spanning trees based on edge betweenness and edge 
information. 
By using as sample cases the cities of Bologna and San Francisco, 
we show how the obtained trees are radically different from those 
based on edge lengths, and allow an extended comprehension of 
the ``skeleton'' of most important routes that so much affects 
pedestrian/vehicular flows, retail commerce vitality, land-use
separation, urban crime and collective dynamical behaviours.
\end{abstract}
\vspace{0.5cm}
\maketitle

%
Centrality is a fundamental concept in network analysis. 
The issue of structural centrality was introduced in the 40's in the 
context of social systems, where it was assumed a relation between the location 
of an individual in the network and its influence in group 
processes \cite{wasserman94}.  
Since then, various measures have been proposed over the years 
to quantify the importance of nodes and edges of a graph,  
and the concept of centrality have found many applications also in  
biology and technology \cite{bareport,newmanreview,vespignanibook,report}. 

In economic geography as well as in regional planning, centrality has 
been dominating the scene 
especially since the Sixties and Seventies stressing the idea that some 
places (cities, settlements) are more important than others because they 
are more ``accessible'', where accessibility was intended as a centrality 
measure of the same kind of those developed in the field of structural 
sociology, with the difference that the geographic nature 
of elements in space was saved around a notion of metric distance \cite{wilson2000}. 
In the field of urban design, a long-term effort has been spent in order to 
understand what urban streets and routes would constitute the ``skeleton'' of 
a city, which means the chains of urban spaces that are most important for 
both the connectedness, liveability and safety at the local scale 
\cite{hillier84,hillier96} and its legibility in terms of 
human wayfinding \cite{burgess99}; more recently, these latter two 
approaches are seemingly merging together in the first clues of a 
cognitive/configurational theory \cite{penn03}. 
After an in-depth investigation of both the topological (dual) \cite{porta_epb1} 
and spatial (primal) \cite{portacentrality,porta_epb2} graph representation 
of street networks, in this paper we provide a tool for the 
analysis of the backbone of a complex urban system represented as a  
spatial (planar) graph. Such a tool is based on the mathematical concept 
of spanning trees, and on the efficiency of centrality measures in 
capturing the essential edges of a graph. 
Differently from previous applications of this same concept \cite{tree_bet}, 
we consider spatial networks instead of 
topological ones, so that our trees can be shown graphically on the 
city maps and can serve as a support in urban design and planning; 
moreover, we consider two different kinds of edge centrality measures, 
and we compare the obtained trees with the standard spanning 
trees based on minimizing the total lengths.

\begin{figure}[htb]
\includegraphics[width=0.47\textwidth]{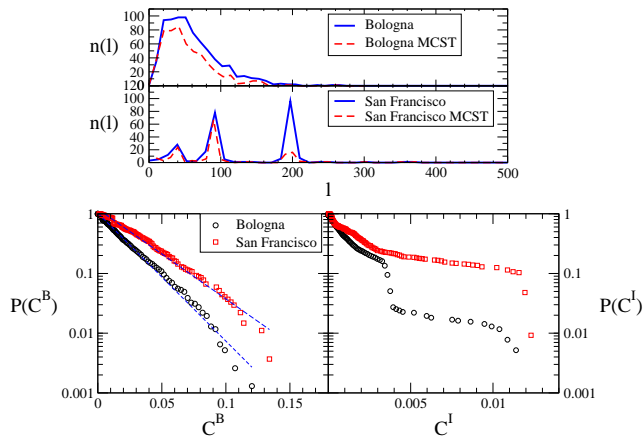}  
\caption{Top panels: the length distributions for the two cities of 
Bologna and San Francisco (full lines) are compared with the length distributions 
of the respective betweenness-based MCSTs (dashed lines). The quantity 
$n(l)$ is defined as the number of edges whose length is in the range 
[l - 5 meters, l + 5 meters]. 
Bottom panels: cumulative distributions of edge betweenness $C^B$ (left)  
and information $C^I$ (right) for Bologna (circles) and San 
Francisco (squares). The dashed lines in the left panel are 
exponential fits to the betweenness distributions. 
}
\label{distributions} 
\end{figure}

%
In our approach, cities are represented as spatial networks 
(networks embedded in the real space), i.e. networks whose nodes occupy a 
precise position in a two-dimensional Euclidean space, and whose edges are real physical 
connections \cite{report,west95}. 
In such  approach, 1-square mile samples of urban street patterns 
selected from Ref.~\cite{jacobs} are transformed into spatial undirected 
graphs by mapping the intersections into the graph nodes and the roads into  
links between nodes \cite{portacentrality,porta_epb2}. Here 
we will focus, in particular, on the cities of {\em Bologna} and 
{\em San Francisco} as examples of two different classes of urban street 
patterns, the former being a {\em self-organized} organic network evolved 
over a long period of time through the uncoordinated contribution of countless 
historical agents while the latter being a mostly 
{\em planned} fabric built in a relatively short period of time following the 
ideas of one coordinating historical agent. 
Each of the two obtained graphs is denoted as $G \equiv G(N,K)$, 
where $N$ and $K$ are, respectively, 
the number of nodes and links in the graph. 
In the case of Bologna we have $N=541$ and $K=773$, while in the case of 
San Francisco the same amount of 1-square mile of land contains only $N=169$ 
and $K=271$ edges. The average degree $<k> =2K/N$ is respectively equal to 
2.71 and 3.21. This difference is due to the overbundance of three-roads 
intersections with respect to four-roads intersections in the city of Bologna. 
The converse is true for the city of San Francisco, due to its square-grid 
structure. See Ref.~\cite{cardillo} for a plot of the entire degree 
distributions in the two cases. 
The graph nodes are characterized by their positions in the unit square 
$\{x_i,y_i\}_{i=1,...,N}$, while the links follow the footprints 
of real streets and are associated a set of real positive 
numbers representing the street lengths, $\{l_{\alpha}\}_{\alpha=1,...,K}$. 
Another relevant difference between the two cities is captured by 
the edges length distribution. In Fig.~\ref{distributions} we plot $n(l)$, 
the number of edges of length $l$, as a function of $l$.  
The edges length distribution has a single peak in Bologna, 
while it has more than one peak in a  mostly planned cities 
as San Francisco, due to its grid pattern. 
In the following, the graph representing a city 
is described by the adjacency $N \times N$ matrix $A$, whose 
entry $a_{ij}$ is equal to 1 when there is an edge between $i$ and $j$ and 
0 otherwise, and by a $N \times N$ matrix $L$, whose entry $l_{ij}$ is the 
value associated to the edge $\alpha \equiv (i,j)$, 
in our case the metric length of the street connecting $i$ and $j$. 
%
%

%
In a previous work \cite{portacentrality}, different measures of 
{\em node centrality} \cite{centrality,lm05a}, properly extended for 
spatial graphs, have been investigated in the same database of urban street patterns. 
Here we show how to construct spanning trees based on {\em edge centrality}.    
We first localize high centrality edges, namely the 
streets that are structurally made to be traversed 
(betweenness centrality) or the streets whose deactivation  
affects the global properties of the system (information centrality). 
Of course other definitions of edge centrality (as for 
instance range, closeness or straightness \cite{centrality})  
can be used as well.  
The definitions of edge betweenness and edge information we adopt are 
obvious modifications of the centrality measures defined on nodes. 

\noindent 
The {\em edge betweenness centrality}, $C^B$, is based on the idea that an edge 
is central if it is included in many of the shortest paths connecting couples 
of nodes. The betweenness centrality of edge $\alpha=1,...,K$ is 
defined as \cite{ng04}: 
\begin{equation} 
 \label{BC}
    C^B_{\alpha}  = \frac{1}{(N-1)(N-2)} 
    \sum_{j, k=1,..,N; j \neq k \neq i} 
    \frac{n_{jk}(\alpha)}{ n_{jk}}  
\end{equation}
where $n_{jk}$ is the number of shortest paths between nodes $j$ and $k$, 
and $n_{jk} (\alpha)$ is the number of shortest paths between nodes $j$ and
$k$ that contain edge $\alpha$.

\noindent
The {\em edge information centrality}, $C^I$, is a measure relating 
the edge importance to the ability of the 
network to respond to the deactivation of the edge itself. The network performance, 
before and after a certain edge is deactivated, is measured by the efficiency of the 
graph $G$ \cite{lm01,lm03}. The information centrality of edge $\alpha$ is defined 
as the relative drop in the network efficiency caused by the removal from $G$ of the 
edge $\alpha$ \cite{portacentrality,lm05a}:
\begin{equation} 
\label{IC}
C^I_{\alpha}  =  \frac{\Delta E}{E} = 
                \frac{E[G] - E[ G^{\prime}]}{E[ G]}
\end{equation}
where the efficiency of a graph $G$ is defined as: 
\begin{equation} 
\label{efficiency}
  E[ G]  =  \frac{1}{N(N-1)} { {\sum_{{i, j=1,..,N; i \neq j }}}  
\frac{d^{Eucl}_{ij}}{ d_{ij} } } 
\end{equation}
and where $G^{\prime}$  is the graph with $N$ nodes and $K-1$ edges
obtained by removing edge $\alpha$ from the original graph $G$. 
An advantange of using the efficiency instead of the characteristic 
path length $L$ \cite{ws98} to measure the performance of a graph is 
that $E[G]$ is finite even for disconnected graphs.

In Fig.~\ref{distributions} we report the cumulative 
distributions of edge betweenness and information. 
The cumulative distribution $P(C)$ is defined as: 
\begin{equation}
  P(C) = \int_C^{+\infty} \frac{n(C^{\prime})}{K} dC^{\prime} 
\end{equation}  
where $n(C)$ is the number of edges with centrality equal 
to $C$. 
The edge distributions are quite similar in the two cities 
of Bologna and San Francisco.  
In particular, the betweenness distributions are well fitted 
by exponential curves, $P(C^B) \sim exp(-C^B/s)$,  
with coefficients respectively equal to $s_{Bo} = 0.020$ and  
$s_{SF} = 0.029$. Thus, for the edge betweenness, the distributions 
found are similar (single-scale) to those observed for the node 
betweenness.  
Conversely, the edge information distributions have not a well 
defined shape: although their decay is slower than exponential 
in both Bologna and San Francisco, the edge information distributions 
do not allow to differentiate 
self-organized cities from planned ones, as it was instead possible 
by means of the node information distributions \cite{portacentrality}. 
This indicates that there are important correlations in the information 
centrality of edges incident in the same node. This also indicates 
that organic self-organized cities are different from planned ones, 
more in terms of their nodes (intersections) than of their edges 
(streets), and especially about how they assign importance to 
such spaces.

%
We are finally ready to build the {\em Maximum Centrality Spanning Trees (MCSTs)}, 
i.e. maximum weight spanning trees where the edge weight is defined as the  
centrality of the edge. 
A graph $G^{\prime} (N^{\prime}, K^{\prime})$ is a {\em tree} if and only 
if it satisfies 
{\em any} of the following four conditions: 
1) $G^{\prime}$ has $N^{\prime}-1$ edges and no cycles; 
2) $G^{\prime}$ has $N^{\prime}-1$ edges and is connected; 
3) exactly one simple path connects each pair of nodes in $G^{\prime}$;
4) $G^{\prime}$ is connected, but removing any edge disconnects it.
Given a connected, undirected graph $G(N,K)$, a {\em spanning tree} $T$ is a 
subgraph of $G$ which is a tree and connects all the $N$ nodes together. 
Consequently $T \equiv  T(N,N-1)$. 
A single graph can have many different spanning trees. 
We can also assign a weight $w_{\alpha}$ to each edge $\alpha$, which 
is usually a number representing how favorable (for instance how central) 
the edge is, and assign a weight to a spanning tree by computing the sum 
of the weights of the edges in that spanning tree. 
A maximum weight spanning tree is then a spanning 
tree with weight larger than or equal to the weight of every other spanning tree 
of the graph. It appears evident that it is possible to define appropriate edge 
weights with the aim of finding particular structures capable of connecting every 
single node of the graph while minimizing the corresponding total weight. 
In particular, for each city we have computed two different MCSTs, 
respectively based on betwenness and information. 
The two cases are obtained by respectively fixing $w_{\alpha} = C^B_{\alpha}$ 
and $w_{\alpha} = C^I_{\alpha}$, with $\alpha = 1,...,K$.  
Since the two centrality measures focus on different properties of the network, 
using both of them allows us to enforce our analysis. 
%
%
\begin{figure}[htb]
\includegraphics[width=0.47\textwidth]{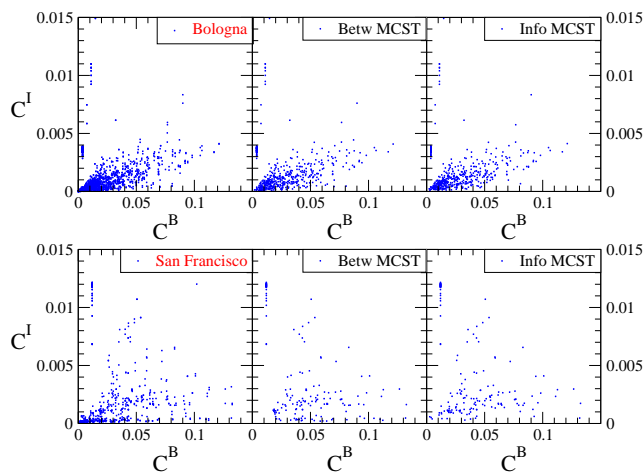}  
\caption{Scatter plots showing the correlations between edge betweenness and 
edge information in Bologna (top panels) and San Francisco (bottom panels). 
Each point represents one edge in the orginal graph (left), in the 
betweenness-based MCST (center), and in the information-based MCST (right). 
}
\label{correlations} 
\end{figure}
Moreover, as shown in Fig.~\ref{correlations} left panels, $C^B$ and $C^I$ 
are correlated, although it is possible to find edges with a low value 
of $C^B$ and a high $C^I$ (and viceversa). 
The coefficients of linear correlation are respectively equal to 
$r=0.69$ and $r=0.46$. 
For the computation of the MCSTs (and of the mLSTs) we have used the Prim's 
algorithm \cite{cormen} that allows to obtain the result in a time 
proportional to $K\log N$. 
The MCST for the city of Bologna contains $K'=N-1=540$ links, i.e. 
$70\%$ of the links of the original graph, while the MCST for San Francisco 
has  $K'=168$, i.e. $62\%$ of the links of the original graph. 
Since the links have been chosen according to their centrality values, 
it turns out that the set of selected edges in the betweenness-based 
MCST of Bologna (San Francisco) possesses the $86\%$ ( $82\%$) of the 
total betweenness centrality of the original graph,  
defined as $\sum_{\alpha=1,K} C^B_{\alpha}$ \cite{tree_bet}. 
Similarly, the set of selected edges in the information-based 
MCST of Bologna (San  Francisco) possesses the $84\%$ ($95\%$) of the 
total information centrality. This is both due to the shapes of the 
centrality distributions shown in Fig.~\ref{distributions} and to the 
edge selection that avoids, in the tree construction, the formation of 
cycles. 
The values of $C^B$ and $C^I$ for the selected edges are shown in the 
scatter plots of Fig.~\ref{correlations}. In the case 
of Bologna, the two measures of centrality have the same correlations 
as in the original graph (the correlation coefficients in the MCST are  
$r^B=0.61$ and  $r^I=0.64$). Conversely, in San Francisco,  
the two variables are less correlated in the MCSTs  
($r^B=0.10$ and  $r^I=0.29$) than in the original graph ($r=0.46$). 
In Fig.~\ref{distributions} (top panels) we have plotted the edge length 
distributions of the betweenness-based MCSTs (dashed lines). 
It is interesting to observe that, for the city of Bologna,  
$n(l)$ has the same shape both in the original graph and in its 
betweenness-based MCST. This means that, in the construction of the 
tree, edges with all lengths have been removed (with same probability) 
from the original graph. 
Conversely, in San Francisco most of the edges not included  
in the betweenness-based MCST are those with the largest 
length. The same result has been found for the information-based MCSTs 
and seems to be a common characteristic of other planned grid-like 
cities. 

\begin{figure*}[htb]
\includegraphics[width=\textwidth]{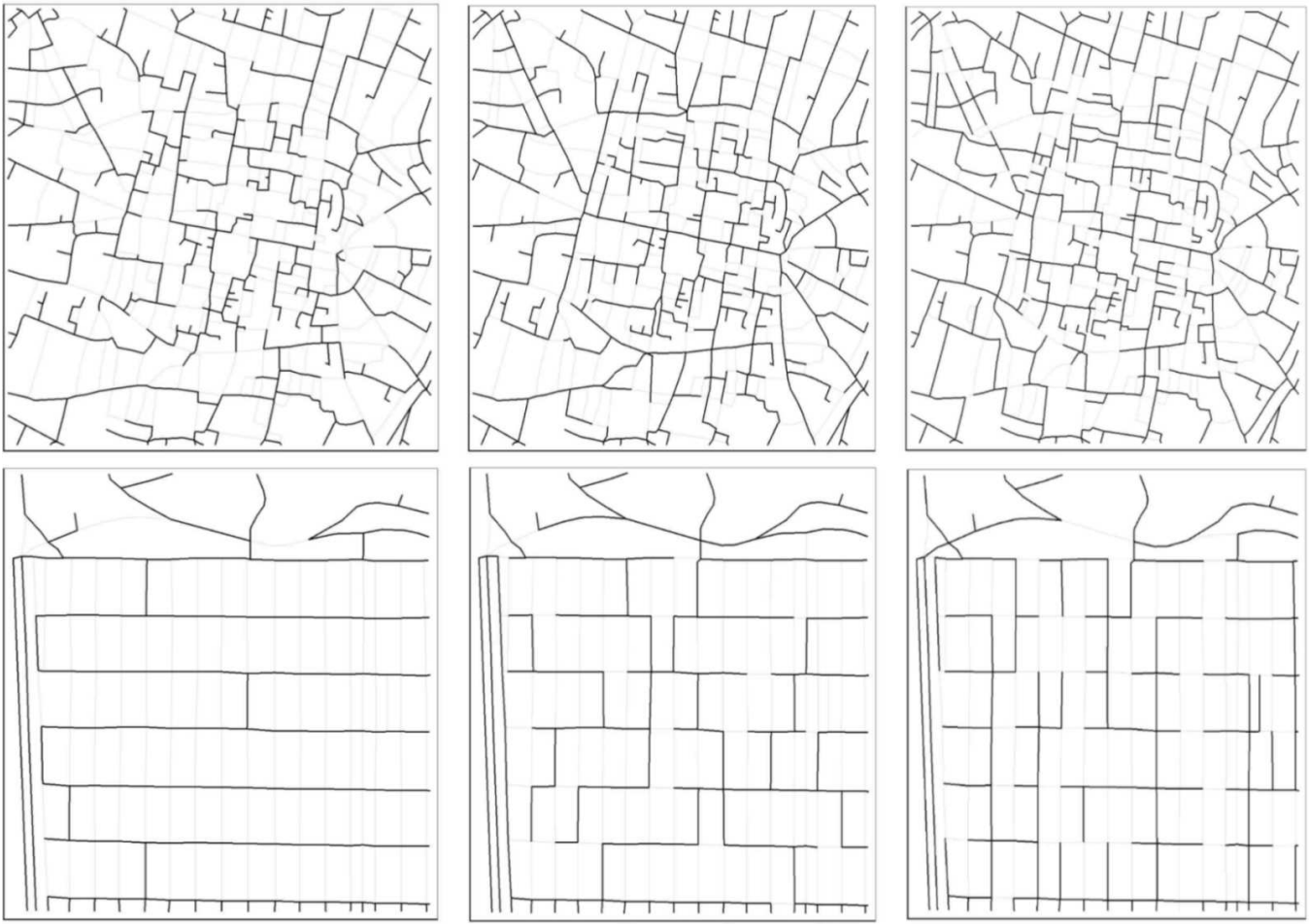}  
\caption{Spanning trees of Bologna (above) and San Francisco (below).  
From left to right, mLSTs, betweenness-based and information-based MCSTs}
\label{MSTs} 
\end{figure*}

In Fig.\ref{MSTs} we compare graphically the two MCSTs with the minimum 
length spanning trees \cite{cormen}.   
In the construction of the latter, the weight $w_{\alpha}$ associated to 
each edge $\alpha$ is set to be equal to the length of the edge $l_{\alpha}$ 
and represent the cost of the edge. 
A {\em Minimum Length Spanning Tree (mLST)} is then a spanning 
tree with weight (cost) smaller than or equal to the weight 
of every other spanning tree of the graph. 
%
%
The MCSTs obtained are different from the mLSTs. In the case of Bologna, the  
betweenness (information) based MCST has a total length equal to 1.15 (1.14) 
times the total length of the mLST, while in the case of San Francisco this 
ratio is equal to 1.15 (1.07). 
In the case of Bologna, the MCST based on betweenness (information) 
has $82\%$ ($75\%$) of the edges in common with the mLST, while in 
San Francisco it has $70\%$ ($76\%$) of the edges in common with the mLST. 
It is worth noting that the two MCSTs have $77\%$ of the edges in common 
in Bologna, whereas such a percentage is smaller in San Francisco ($66\%$). 
The graphical visualization of the maximum centrality trees is of interest 
for urban planners since the trees express the uninterrupted 
chain of urban spaces that serves the whole system 
while maximizing centrality over all edges involved. This method 
identifies the backbone of a city system as the sub-network of spaces 
that are most likely to offer the highest potential for the life of 
the urban community in terms of popularity, 
safety and services locations, all factors geographically related with 
central places. 
This is evident in Fig.~\ref{MSTs}, where the comparison between the trees 
in the two cities clearly indicates that the spatial sub-system 
that keeps together a city in terms of the shortest trip length is not 
the same spatial sub-system that does it in terms of the highest centrality. 
It is also worth noting that metric distance is also involved in 
the algorithms for the calculation of centrality indices, so 
that all kinds of trees considered hereby are rooted in the geographic 
space. 
The second thing is that while the shortest length backbone performs 
effectively when applied to {\em planned} urban fabrics like San Francisco, 
in {\em self-organized} evolutionary cases like that of Bologna it does 
not individuates continuous routes nor clearly distinguishes 
a hierarchy of sub-systems in the network, while the highest 
information and especially the highest betweenness backbones do. 
In a way, we would say that organic patterns are more oriented to put 
things and people together in public space than to shorten the trips from 
any origin to any destination in the system, this latter character being 
more typical of planned cities.

In conclusion, in this work we have shown that the concept of MCST 
leads to a meaningful picture of the primary sub-system of a city network, 
which makes it a single component while minimizing the cost of moving around and 
maximizing the potential of places to achieve social success, safety and 
popularity. Therefore, the method has the potential of becoming an useful 
tool in city planning and design, due to its immediate and powerful 
visualization outcome.

{\bf Acknowledgment.} 
\noindent
We thank P. Crucitti for many helpful discussions and suggestions.

\end{document}